\documentclass[final] {aipproc}

\layoutstyle{6x9}

\SetInternalRegister\hbadness{8000} 

\begin{document}

\title 
      {Conference on the Intersections of Particle and Nuclear Physics 2003: \newline  Relativistic Heavy Ion Parallel Session Summary}

\classification{43.35.Ei, 78.60.Mq}
\keywords{Document processing, Class file writing, \LaTeXe{}}

\author{J.L. Nagle}{
  address={University of Colorado at Boulder},
  email={jamie.nagle@colorado.edu},
}

\author{T. Hallman}{
  address={Brookhaven National Laboratory},
  email={hallman@bnl.gov},
}

\copyrightyear  {2003}

\begin{abstract}
The Relativistic Heavy Ion Collider (RHIC) came online in 2000, and the last three years have
provided a wealth of new experimental data and theoretical work in this new energy frontier for nuclear
physics.  The transition from quarks and gluons bound into hadrons to a deconfined quark-gluon
plasma is expected to occur at these energies, and the effort to understand the time evolution of these
complex systems has been significantly advanced.  The heavy ion parallel session talks from the Conference 
on the Intersections of Particle and Nuclear Physics (CIPANP) 2003 are posted at:
``http://www.phenix.bnl.gov/WWW/publish/nagle/CIPANP/''.  We provide a brief summary of these sessions
here.
\end{abstract}

\date{\today}

\maketitle

\section{Introduction}

Many speakers separated the topics of heavy ion reactions into two broad categories.  (1) The bulk 
system created which includes of order 5000 hadrons - with transverse momentum below 1.0 GeV/c.  These
hadrons result from the interaction of approximately 10,000 virtual gluons liberated from the nuclear wavefunction upon impact.  
These particles contain information about equilibration and thus the equation of state.  (2) The probes of the bulk system, that
are defined as colored systems such as hard scattered quarks and gluons that traverse the bulk media and 
tell us about its properties.  Although these high $p_{T}$ partons or heavy quark states are relatively rare, and
thus do not impact the equation of state (they are very non-equilibrium), because they are accessible via
pQCD or QCD phenomenology, they represent a key calibration of the system properties.  M. Gyulassy referred
to the importance of these pQCD probes as analogous to the ``tail that wags the QGP dog.''

\section{Bulk Effects}

P. Steinberg reviewed the contrasting pictures of bulk thermalization and initial state scaling behavior.  The former 
implies a collective system whose properties are dictated by the dynamics after the first collision impact, while the latter are
remnants of the initial parton distributions or first reaction mechanisms.
P. Kolb pointed out that it appears for the first time in nuclear reactions, that hydrodynamics with 
no additional viscosity or multi-fluid
expressions appears to describe the bulk momentum distribution of hadrons at low $p_{T}$,
including the azimuthal anisotropy.  However,
the calculations appear to yield too long a lifetime for the system relative to two particle correlation
measurements (HBT) or $R_{out}$.  The calculations indicate enormous initial pressure, with a system
decoupling rather quickly, of the time scale of 10 fm/c.  P. Kolb gave a status report on the success of these
calculations, and that some require a phase transition to extend the lifetime thus giving additional boost to
the heavier particles.

M. Lisa gave a detailed presentation of the nearly complete experimental information at central rapidity on the phase space 
distribution - both momentum and position - for
hadrons at the freeze-out (point of final interaction).  Good progress has been made toward tracing back
in time to the initial conditions and the equation of state.  D. Magestro showed the outstanding agreement 
between statistical models and hadron ratios, including strange and multi-strange baryons.  However, the
suggestion that the fast time scale for thermalization implies that the system is 
``born into'' into equilibrium was not a clear physical picture.
Z. Xu and A. Tang presented data on flow and more exotic hadron production that help to complete the hadronic 
freeze-out picture.  Results from STAR indicate modifications to the low mass vector mesons as observed
via hadronic decay channels.  Comparing these with reconstruction in the leptonic channels will be an interesting
future measurement from PHENIX.  R. Debbe showed that the BRAHMS hadron yields at forward rapidity appear as 
part of a ``different source'' than particles in the mid-rapidity region with a different net baryon density.


Y. Kovchegov and K. Tuchin presented global and azimuthal anisotropy observables that were generated with
only initial state effects included - using a parton saturation picture as the calculation tool.  If correct, this
implies that the measured $v_{2}$ has no relation to a physical reaction plane.  M. Lisa pointed out that the
HBT results into and out of the reaction plane seem to disprove this assertion.

\section{Probes}

An excellent probe of the medium is a hard scattered parton (quark or gluon) that transverses the medium.  One
can use QCD factorization to calculate the expected rate of these partons and their resulting fragmentation
hadron products.  It has already been observed by all four RHIC experiments in Au-Au reactions, 
that high $p_{T}$ hadrons are suppressed
relative to this particular expectation.  Since a parton scattering through a dense gluonic medium loses additional
energy via gluon bremsstrahlung, it can be used as a ``gluonometer'' - a calibratable probe of the color charge density
of the medium.  Another explanation for the lack of high $p_{T}$ hadrons is that the high virtual parton density in the
incoming nuclei saturates, and thus distinct partons are replaced by a gluon wavefunction that inherently break
factorization.  This saturation effect that might be probed in protons at $x \approx 10^{-4}$ (for example in HERA DIS), moves
to $x \approx 10^{-2}$ in heavy nuclei due to the additional thickness.  Many attempts to use this framework
- sometimes referred to as the Color Glass Condensate (CGC) - to describe the bulk global observables have shown some success.  More
recently, D. Kharzeev and collaborators tried to extend the saturation regime via DGLAP evolution to higher $Q^{2}$, referred
to as the Color Quantum Liquid (CQL) regime.  This CQL region suppressed hard scattering for hadrons even up to $p_{T} \approx 8$
~GeV/c - thus also providing a possible explanation of suppressed high $p_{T}$ hadrons observed in Au-Au reactions.

This ambiguity led the RHIC experiments to propose a simple plan to resolve the question of whether the
suppression is an initial state or final state effect.  RHIC ran deuteron-Au reactions at the start of Run-3, since
for the CQL picture, the suppression should still be present in the Au nucleus.  In contrast, as pointed out by
I. Vitev and others, the final state energy loss explanation would not be applicable with
no dense medium created in deuteron-Au collisions.

Data from PHENIX, PHOBOS, and STAR as presented by 
L. Aphecetche, G. Roland, and P. Jacbos show an enhancement of high $p_{T}$ hadrons (both 
unidentified and $\pi^{0}$) relative to factorization and nuclear thickness scaling in deuteron-Au reactions - as 
sometimes referred to as the Cronin effect.  There is no observed suppression as predicted in the
saturation model extension.
Angular correlations also revealed very similar near-side and away-side jet
structure as observed in proton-proton reactions.  M. Miller and J. Rak emphasized that the disappearance of away
side correlations in Au-Au reactions, is not seen in deuteron-Au reactions.
Though there was much discussion - including talks by J. Jalilian Marion and L. McLerran - 
there was general agreement that the data rule out the saturation model extension 
- referred to as the Color Quantum Liquid regime, for $x \approx 10^{-1}$.    
This of course does not rule out saturation at lower x having an effect on the bulk dynamics
as pointed out by L. McLerran and others.

Many speakers stated that with the extension to the saturation picture ruled our, parton energy loss in
the dense gluonic medium is the only answer.  L. McLerran stated that ``if a 10 GeV quark loses 2-3 GeV
of energy in medium, it seems clear that the bulk partonic matter is thermalized.''  However, challenging
questions still remain that must be fully explored.  T. Chujo and J. Klay pointed out that although light hadrons
are suppressed, at least at intermediate $p_{T}$, the baryons - both protons and $\Lambda$ are not.  S. Bass and
B. Muller both argued that while very high $p_{T}$ hadrons may be dominated by partons that lose energy in medium
and that fragment into hadrons in vacuum, intermediate $p_{T}$ hadrons may come from parton recombination.  This
picture shows great promise in particular in the explanation of the $v_{2}$ scaling of different hadrons.  However,
the model is simple - not including gluon contributions - and further tests including $\phi$ mesons - large
mass but $q\overline{q}$ state - are needed.

HERMES electron-nucleus deep inelastic scattering (DIS) data indicate modified fragmentation functions as shown
by E. Kinney.  These have been interpreted in the framework of fast hadronization or color dipole
formation followed by re-interaction in the nucleas, and also in terms of multiple scattering from color
charges in the nucleons of the nucleus inducing gluon bremsstrahlung.  
In Au-Au reactions, modification of QCD vacuum may also effect zeroth order gluon radiation, as described by M. Djordvic with 
regards to heavy quarks, and thus may also play a role for light partons.  X.N. Wang argued that only the partonic energy
loss explanation survives, but there was much debate and a well developed model including fast color dipole formation or
hadronization is still lacking - that then needs to be confronted with all data.  There is a wealth of new 
data that needs to be looked at in detail, and new ideas considered for a full description.

\section{Heavy Flavor Probes}

H. Woehre gave an overview of charm measurements at lower energies and detailed the constraints those place
on heavy flavor production at higher energies.  J. Heuser presented the detector performance of NA60, and it
is clear that they have the ability to resolve the low mass vector meson states - down to low $p_{T}$ - and
the $\chi_{c}$ contribution to the $J/\psi$.  Many expressed that it should be a high priority for the field that they get adequate
running time.  First charm result via ``prompt'' single electrons from PHENIX
were shown by S. Batsouli and indicate a surprising feature. The spectra agree both with a unmodified charm
quark fragmentation in vacuum and with a completely thermalized charm plus a hydrodynamic boost.  

D. Silvermyr showed the first $J/\psi$ results from at RHIC from PHENIX in proton-proton and Au-Au reactions.
Despite low statistics, the Au-Au results appear to disfavor models of large $J/\psi$ enhancement relative
to binary scaling.  However, future high statistics measurements are critical to utilitize these
quarkonia probes.
The $J/\psi$ measurement in proton-proton collisions
 of the total cross section and mean $p_{T}$ are the first at collider energies and should help
address color-singlet versus color-octet production mechanisms.

\section{Summary of the Summary}

The heavy ion sessions included the presentation of a wealth of new data on high $p_{T}$ probes, including
the first deuteron-Au results.  It is now clear that the suppression of high $p_{T}$ hadrons is the
result of a final state effect or some complete break-down of factorization - but not in the initial
wavefunction.  There was lively discussion of whether this is definitely partonic multiple scattering
in the colored medium, or if contributions from fast hadronization, recombination, or zeroth order
radiation modification may still play a role.  New results on intermediate $p_{T}$ baryons, 
low mass vector mesons, charm mesons,
and quarkonia were shown and more data from these states will help to fill in the picture.  For the bulk
medium an overwhelming amount of data already exists and must be reconciled in terms of hydrodynamic pictures and
observations of simple scalings.  

We would like to thank all the speakers for their excellent presentations and all participants for
spirited discussions.


\begin{theacknowledgments}
We would like to thank the organizers of the conference - Dr. W.J. Marciano, Dr. Z. Parsa and the organizing committee -
for all their hard work in coordinating this event.  JLN acknowledges support by the U.S. Department of Energy under Grant No. 
DE-FG02-00ER41152 and the Alfred P. Sloan Foundation.
\end{theacknowledgments}



\end{document}